\documentclass[12pt]{article}
\usepackage{amsmath}
\usepackage{amsfonts}

\newcommand{\vet}[1]{{#1}}
\newcommand {\beq} {\begin{equation}}
\newcommand {\eeq} {\end{equation}}
 \newcommand {\ber}{\begin{eqnarray*}}
 \newcommand {\eer} {\end{eqnarray*}}
\newcommand {\bea}{\begin{eqnarray}}
 \newcommand {\eea} {\end{eqnarray}}

\def\one{{\hbox{ 1\kern-.8mm l}}}
\newcommand{\Tr}{{\rm Tr}}
\newcommand{\sect}[1]{\setcounter{equation}{0}\section{#1}}
\renewcommand{\theequation}{\thesection.\arabic{equation}}

\def\a{\alpha}

\def\e{\epsilon}

\def\l{\lambda} 
\def\m{\mu}

\def\th{\theta}

\def\ii{{\rm i}}
\newcommand{\ex}[1]{{\rm e}^{#1}} \def\ii{{\rm i}}

\newcommand{\be}{\begin{equation}}
\newcommand{\ee}{\end{equation}}
\newcommand{\eq}[1]{(\ref{#1})}

\begin{document}
\linespread{1.2}

\hfill{LPTENS-03/22}

\hfill{DFTT 14/03}

\vspace{20pt}

\begin{center}

{\LARGE \bf Twisted determinants on higher genus Riemann surfaces}

\vspace{30pt}

{\bf Rodolfo Russo$^{a}$ and Stefano Sciuto$^{b}$ }

\vspace{15pt}
{\small \em
\begin{itemize}
\item[$^a$]
Laboratoire de Physique Th\'eorique de
l'Ecole Normale Sup\'erieure,  \\
24 rue Lhomond, {}F-75231 Paris Cedex 05, France
\item[$^b$]
Dipartimento di Fisica Teorica, Universit\`a di Torino;\\
I.N.F.N., Sezione di Torino, Via P. Giuria 1,
I-10125 Torino, Italy
\end{itemize}
}

\vskip .1in {\small \sffamily
rodolfo.russo@lpt.ens.fr, sciuto@to.infn.it}
\vspace{50pt}

{\bf Abstract}
\end{center}

\noindent
We study the Dirac and the Laplacian operators on orientable Riemann
surfaces of arbitrary genus $g$. In particular we compute their
determinants with twisted boundary conditions along the
$b$--cycles. All the ingredients of the final results (including the
normalizations) are explicitly written in terms of the Schottky
parametrization of the Riemann surface. By using the bosonization
equivalence, we derive a multi-loop generalization of the well-known
$g=1$ product formulae for the Theta--functions. We finally comment on
the applications of these results to the perturbative theory of open
charged strings.

\setcounter{page}0 \setcounter{footnote}0 \thispagestyle{empty}
\newpage
\linespread{1}

\sect{Introduction}

The perturbative expansion of string theory has been thoroughly
studied in the last thirty years and the first works date back to the
time of the dual models~\cite{Alessandrini:1971dd}. Our understanding
of this subject has greatly increased during the years, in particular
in the eighties, with many conceptual and technical breakthroughs (see
for instance~\cite{Polyakov:rd}). The geometrical meaning of string
amplitudes became much more clear and the strict relation of these
physical quantities with the theory of Riemann surfaces and
Theta--functions made the subject even more interesting.

Among the basic building blocks of string amplitudes one finds the
determinants of the Laplacian and of the Dirac operator.
Since these operators are present in the 2D action describing the free
string propagation, their determinants will appear in all perturbative
amplitudes, starting from the simplest one: the vacuum energy. A
detailed study of these determinants was performed in the
papers~\cite{Knizhnik:kf,Alvarez-Gaume:1986es,Verlinde:1986kw,Alvarez-Gaume:1987vm,Eguchi:1986ui,Lechtenfeld:1989wu}.

It is particularly interesting to compare the fermionic and the
bosonic results, because, even though they have a very different
structure, they are related through the bosonization duality. Thus the
equivalence between the fermionic and the bosonic determinants can be
used as a tool for giving ``physicist proofs'' of interesting
mathematical identities~\cite{fay1,mumf,fay2}. In fact it is known
that from general addition theorems for Theta--functions (see, for
example, Proposition~2.16 of~\cite{fay1}) one can derive some
identities proving the equivalence between bosonic and fermionic
systems on a $g$--loop Riemann surface. For instance, the identity in
Corollary~2.17 of~\cite{fay1} is a consequence (or a proof, according
to the points of view) of the relation between a fermionic
system\footnote{See Appendix A for the definition of our conventions.}
$(b,c)$ of spin $(1,0)$ and a chiral boson of background charge $Q=-1$:
\beq\label{2.17}
\th\left(\sum_{i=1}^g J(z_i)-J(w)-\Delta\Big|\,\tau \right)
\frac{\prod_{i<j=1}^g E(z_i,z_j)}{\prod_{i=1}^g E(z_i,w)}
\frac{\prod_{i=1}^g\sigma(z_i)}{\sigma(w)} =
C~{\rm det} \left[\omega_i(z_j)\right]~.
\eeq
In the mathematical literature, the analysis of identities like the
one above focuses on the dependence of the various functions on the
punctures $z_i$. Thus often these equalities are written in terms of
some ``constants'' (like $C$ in the above equation) that depend only
on the moduli of the surface, but not the punctures $z_i$. However, it
is also quite interesting to give an explicit expressions of these
constants, since they are closely related to the partition functions
of the bosonic and fermionic systems. A powerful technique for writing
explicitly $C$ in terms of the moduli of the surface is the sewing
technique. This is a very old idea~\cite{Alessandrini:1971dd} allowing
to construct higher loop amplitudes from tree diagrams: pairs of
external legs are sewn together taking the trace over all possible
states with the insertion of a propagator that geometrically
identifies the neighborhoods around two punctures. The results
obtained in this way give rise to a particular parametrization of the
$g$--loop Riemann surface known as Schottky uniformization. At
$1$--loop, string theorists are very familiar with this
parametrization: in this case, the torus is just seen as the complex
plane where two points $z,\,w$ are identified if $z = k^n w$, $\forall
n\in{\mathbb Z}$. Here $k$ is a complex parameter representing the modulus of
the torus and is related to period matrix entering in the
Theta--functions by $k=\exp{(2\pi\ii \tau)}$, with ${\rm Im} \tau > 0
\Leftrightarrow |k| <1$. At $1$--loop the constant $C$ in
Eq.~\eq{2.17} is just $C=\prod_{n=1}^\infty \left(1 -k^n \right)^3$
and is related to the partition function of a chiral scalar
field. This pattern can be generalized to higher genus surfaces by
sewing other handles to the $1$--loop result and one gets
$C=\prod'_\a\prod_{n=1}^\infty \left(1 - k^n_\a \right)^3$, where the
exact meaning of $k_\a$ and of product over the Schottky group
$\prod'_\a$ will be explained later (see Eq.~\eq{tdetH} and
Appendix~B). Here we just want to stress that identities like
Eq.~\eq{2.17} can be exploited also to rewrite products over the
Schottky group in terms of more geometrical quantities like
Theta--function, Abelian differentials, and Prime form (see
Appendix~A).

At $1$--loop, the possibility to pass from the formulation in terms of
Schottky products to the one with Theta--functions has been heavily
exploited in string theory. The physical reason is clear: the
geometrical expressions in terms of Theta--functions has the advantage
to make manifest the modular properties of the string results. This is
in contrast to the expressions written in terms of Schottky products,
where these properties are not manifest. In fact a modular
transformation can be non--analytic in $k$ (for instance, consider
$\tau\to -1/\tau$). This limitation is precisely the reason why the
Schottky uniformization has not been much studied by
mathematicians. However, the Schottky uniformization has an important
physical significance because it makes manifest the unitarity
properties of string amplitudes. This can be easily understood by
remembering that this uniformization naturally arises from the sewing
procedure where the Schottky multipliers $k$ are basically the
equivalent of the (exponential of the) Schwinger parameters in field
theory. Thus the Taylor expansion in the $k$'s of the string results
isolates the contribution to the amplitude coming from the propagation
of particular states in the various handles of the surface. At
$1$--loop level, the relation between $k$ and $\tau$ is particularly
simple and thus one can use also the Theta--function expressions to
analyze the unitarity properties of the string results. However, on a
$g$--loop surface there is no equivalent of the simple relation
$k=\exp{(2\pi\ii \tau)}$ and only the expressions in terms of Schottky
parameters display the unitarity properties in a simple way.

Thus the expressions of string results \`a la Schottky and the one in
terms of Theta--functions capture two different but equally important
features of string theory. Depending on the question one would like to
answer it is more convenient to write the amplitudes in one or the
other form. Because of this, it is quite crucial to be able to rewrite
a general Schottky product in terms of geometrical objects and
vice--versa. As we already said, this step is by now standard in
$1$--loop computations, where the period matrix is just a complex
number which is related in a very simple way to the only Schottky
multiplier $k$. In this case, string theorists often make use of
identities like the one in Eq.~\eq{t11-l}. At mathematical level,
these $1$--loop identity can be proved by showing that both sides of
the equation have the same periodicity, poles and zeros. The
generalization of this kind of identities to the higher loop case is
much harder. However, as we noticed before, it is precisely in the
multiloop expressions where one really needs to have these identities
to pass from a manifest unitarity amplitude to a modular covariant
one.

Thus the main purpose of this paper is to find new relations of the
kind of Eq.~\eq{2.17} that relate Theta--functions to the Schottky
products contained in the constant $C$. Following the ideas
of~\cite{Pezzella:1988jr,Losev:1989fe,DiVecchia:ht}, we use the
equivalence between bosonic and fermionic systems as a device allowing
to recast Schottky products in terms of Theta--functions. We basically
generalize previous
analysis~\cite{Pezzella:1988jr,Losev:1989fe,DiVecchia:ht} in two
directions. On one hand we consider fermions of spin $(\lambda,
1-\lambda)$ and the dual bosonic system with the background charge
$Q=1-2\l$, instead of focusing just on the simplest case
$\lambda=1/2$. On the other hand, for general $\lambda$ we also
consider twisted periodicity conditions along the $b$--cycles. This
twisting can be equivalently thought as the effect of a flat gauge
connection along the $b$--cycles\footnote{As usual, we call
$b$--cycles the loops in the worldsheet along the $\tau$ direction and
$a$--cycles the spatial loops along $\sigma$.} on a minimally coupled
fermionic system. Thus the periodicity parameters ($\e_\mu$) can be
naturally thought as non--geometrical parameters of the Riemann
surface. In fact, in the generalization of Eq.~\eq{2.17} arising in
these cases, the constants $C$ depend also on $\e_\mu$ beyond the
usual dependence on the geometrical moduli. Our goal is to give an
explicit expressions in terms of Schottky series or products for all
the quantities appearing in the various identities. In this way, at
least in principle, one can compute all the ingredients in terms of
the parameters of the surfaces with an arbitrary degree of precision.

One more clarification is due at this point, since it may appear
unclear why in our study $b$--cycles twists have a privileged status
in comparison with $a$--cycle twists. The difference is simply due to
the Hamiltonian approach adopted here, where the twists along the time
direction can be taken into account simply by a modification of the
sewing propagator, while those along the spatial directions modifies
radically the spectrum of the free theory. In the higher genus
diagrams, this difference implies that $b$--cycles twists can be
described \`a la Schottky by means of the usual representations of the
projective group, while the direct description of amplitudes with
$a$--cycle twists seems to require a more complicated formalism (for
an explicit $1$--loop example see \cite{Chu:2002nd} and compare
Eqs.~(7) and (8) therein). Of course, once a result with $b$--cycles
twists is known in terms of Theta--function, one can explicitly
perform a modular transformation and derive the equivalent quantity
with $a$--cycles twists. This strategy can be used to obtain
the partition function of the charged open bosonic string in presence
of a constant external field.

The structure of the paper is the following. In Section~2 we recall
the main features of the sewing technique, focusing on the fermionic
correlators. The presence of general twisted periodicity along the
$b$--cycles requires a modification of the sewing procedure and it is
crucial for the consistency of the results to carefully follow all the
effects due to the presence of $\e_\mu$. The main result of this
section is the explicit expression~\eq{tda} of the twisted abelian
differentials. In Section~3 we compare the fermionic correlators with
the corresponding bosonic ones and derive Eqs.~\eq{t2.17}--\eq{Ce}
which generalize \eq{2.17} to the twisted case.  Finally in the
Conclusions we discuss some applications of our results in the context
of string theory.  Then in Appendix~A we define some quantities of
interest in the theory of Riemann surfaces, in Appendix~B we discuss
the Schottky parametrization and in Appendix~C we give some details of
the sewing technique.

\sect{The sewing technique}

The use of the sewing technique for computing multiloop amplitudes is
discussed in detail in~\cite{DiVecchia:1988cy} for bosonic systems and
in~\cite{DiVecchia:1989id} for fermionic systems. In order to
generalize their results to the twisted case, we will follow
closely these two papers and refer to them for the explicit derivation
of the results used here as starting point. In this section, we just
present the main ideas of the sewing technique in order to clarify its
application to the physical/mathematical problem discussed in this
paper.

\subsection{The torus}
As anticipated in the introduction, the main idea of the sewing
technique is to employ a bootstrap approach and construct $g$--loop
amplitudes starting from the tree-level results. The first step is of
course the construction of $1$-loop amplitudes. 
This derivation is quite natural and well-known since it is explained
in Chapter VIII of Green, Schwarz, and Witten book~\cite{Green:mn}.
In fact, tree-level amplitudes can be written with a special choice
for the puncture of the external states putting the first vertex in
$z=\infty$ and the last in $z=0$. Thus the first state just describes
an outgoing string with $\langle s_1|$ and the last state describes an
ingoing string with the ket $|s_N\rangle$. At this point one can
simply relax the on--shell condition, insert the propagator $k^{L_0}$,
and identify the two states
\beq\label{0-1}
\langle 0| V_1(\infty)V_2(1) \ldots V_N(0)|0\rangle =
\langle s_1| V_2(1) \ldots V_{N-1} |s_N \rangle \to
\Tr\left[V_2(1) \ldots V_{N-1} k^{L_0}\right]~.
\eeq
This transforms the v.e.v. of the vertex operators, typical of
tree-level amplitudes, into a trace over the Hilbert space of a
propagating string; moreover the neighborhood around the points
$z=\infty$ and $z=0$ are identified by the projective transformation
($z\rightarrow k^n z$) generated by the insertion of the propagator.
As an example, it is useful to start with the analysis of the vacuum
$1$--loop amplitude, since it contains some of the features of the
general computation and can be derived from the simple trace
in~\eq{0-1} without external states. In fact, it is not difficult to
compute the torus partition function of a fermionic system of spin
($\l,1-\l$) with trivial periodicity conditions on the $a$--cycle and
twisted ones along the $b$--cycle.
In fact in the sewing construction the $b$--cycles are generated by the
identification of the neighborhoods around two points (usually
$z=\infty$ and $z=0$ at $1$--loop) enforced by the propagator $P$ (one
can take simply $P=k^{L_0}$). Since we want to have non--trivial
periodicity condition along these cycles, we need to deform the
propagator by adding an $\e$ dependence so that
\beq \label{perio}
b(z) = k^{\l}\ex{-2\pi \ii \e}\,\Big(P_\e^{-1} b(k\,z) P_\e\Big)
~~,~~~~
c(z) = k^{1-\l}\ex{2\pi \ii \e}\,\Big(P_\e^{-1} c(k\,z) P_\e\Big)~.
\eeq
This can be obtained simply by inserting together with the usual
propagator $P$ also a factor of $\ex{2\pi\ii \e j_0}$, where $j_0$ is
the fermionic number operator. The torus partition function can
be computed as usual just by taking the trace of twisted $P_\e = P\,
\ex{2\pi\ii \e j_0}$
\beq\label{t1l}
Z^{\l}_\e=\Tr\left[k^{L_0} \ex{2\pi \ii \e j_0}\right]_\lambda =
\prod_{n=\lambda}^\infty \Bigg( \left(1 - \ex{2\pi\ii \e} k^{n}
\right) \left(1 - \ex{-2\pi\ii \e} k^{n} \right)  \Bigg)
\prod_{r=1-\l}^{\l-1} \left(1 - \ex{2\pi\ii \e} k^{r} \right)~.
\eeq
The structure of this result is quite clear: the first two factors
come from the action of the modes $c_{-n}$ and $b_{-n}$ with $n\geq
\l$ respectively, while the last product is related to the special
modes $c_r$ with $r\in[(1-\l),(\l-1)]$ only, since the corresponding
$b_r$ oscillators vanish on the vacuum. If $\l$ is integer the
result~\eq{t1l} can be rewritten as
\beq\label{t1li}
Z^{\l\in \mathbb{N}}_\e = (-1)^\lambda \ex{\ii \pi(2\l -1)\e}
k^{\l(1-\l)/2} \;2 \ii \sin{\pi\e} \prod_{n=1}^\infty
\left(1 - \ex{2\pi\ii \e} k^{n} \right)
\left(1 - \ex{-2\pi\ii \e} k^{n} \right)~,
\eeq
which is clearly vanishing for $\e=0$. This fact has an important
geometrical explanation. For integer $\lambda$, there are periodic and
regular solutions of the equation of motions for both $b$ ($b
\Leftrightarrow z^{-\l}$) and $c$ ($c \Leftrightarrow z^{\l-1}$). In
fact the singularities in $z=0$ an $z=\infty$ are the fixed
points~\eq{fp} of the projective transformation $P=k^{L_0}$ and so are
outside the fundamental region representing the torus generated by the
propagator $P$ (see Appendix~B). In other words, both $b$ and $c$ have
a zero--mode on the torus and thus it is natural that the partition
function without any insertion vanishes. For generic fixed points
$\xi$ and $\eta$ these zero--modes can be written as
\beq\label{zm1l}
b(z) \Leftrightarrow \left[\frac{\eta-\xi}{(z-\eta)(z-\xi)}\right]^\l
~~,~~~~ c(z) \Leftrightarrow
\left[\frac{\eta-\xi}{(z-\eta)(z-\xi)}\right]^{1-\l}~.
\eeq
The presence of a twist along the $b$--cycles lifts both zero--modes,
since the solutions~\eq{zm1l} do not satisfy the new boundary
conditions. This explains why the partition function does not vanish
for $\e\not= 0$. Notice that the difference between the number of the
$b$ zero-modes and the number of the $c$ zero-modes is always zero for
the torus topology, as it should be. In fact the Riemann-Roch theorem
ensures that this difference can depend only on the integrals of the
curvature and of the gauge field strength. As said before, the twists
are equivalent to the presence of a flat connection along the
$b$--cycles, and thus the difference between the number of the $b$ and
the $c$ zero-modes can not depend on $\e$. To be precise, for the
cases we are interested in, the Riemann--Roch theorem says
\beq\label{R-R}
\# c~\mbox{zero--modes} - \# b~\mbox{zero--modes} = Q\, (g-1)~,
\eeq
where, as usual, the background charge $Q$ is related to the spin $\l$
of the fermionic system $(b,c)$: $Q=1-2\l$.

\subsection{Higher genus surfaces}
Turning to the study of higher genus surfaces we observe that this
procedure, however, can not be straightforwardly generalized, because
the vertex operators $V$ in~\eq{0-1} describe just the emission of a
specific on--shell state. They depend only on the quantum numbers of
the emitted string and thus there is no easy way to identify two of
them and sum over all intermediate states. In order to this, one
should use a generalization \cite{Sciuto:1969vz,Neveu:1986mv} of the
vertex operators, where also the emitted states are described by a
whole Hilbert space
\bea \label{V30}
\langle {\cal V}_{I}^X| & =& {}_{I} \langle 0, x_0=0 |
:\exp\left\{\oint_0 dz \left(-X^v(1-z)\partial_z
X_{I}(z)\right)\right\}:~,
\\ \nonumber
\langle {\cal V}_{I}^{bc} | & =& {}_{I}\langle 0; q=-Q|
: \exp\left\{\oint_0 dz \left(b^v(1-z) c_{I}(z) -
c^v(1-z) b_{I}(z) \right) \right\}:~.
\eea
Here the coordinates with the superscript $v$ describe a propagating
(virtual) string, while those with a subscript $I$ describe a generic
emitted state in the Hilbert space ${\cal H}_I$. The modes of the two
types of fields (anti)--commute among them, since they refer to
completely independent states. Let us stress that the bra--vector in
Eq.~\eq{V30} is the vacuum in the Hilbert space of the emitted string,
so that $\langle {\cal V}|$ is an operator in the $v$--Hilbert space
of the propagating string. One can think of the vertices $\langle
{\cal V}|$ as an ``off--shell'' generalization of the usual bosonic
and ghost part (for $Q=-3$) of string vertex operators. However, here
off--shell does not have the usual meaning as in field theory. On the
contrary, off--shell just means that the external states have not been
specified yet, and thus $\langle {\cal V}|$ can be seen as the
generator of all possible three string interaction which are obtained
by saturating it with physical (on--shell) states.

So far, in the presentation, we always made reference to the
space--time interpretation of the formalism typical of string
theory. However, the same formalism can be also applied in the
context of 2--dimensional free conformal field theory. One has just to
consider the two vertices in Eq.~\eq{V30} separately and use them to
construct partition functions or Green functions for bosonic or
fermionic systems.
By using the vertex $\langle {\cal V}^{bc}|$ as ingredient, we can
substitute Eq.~\eq{0-1} with a new formula that is suited for the
generalization of the sewing procedure to higher genus
surfaces. Following~\cite{DiVecchia:1989id} it is easy to construct
the generator of the $N$-point Green function on the sphere ($g=0$)
\beq\label{tree}
\langle V^{bc}_{N;0} | = {}_{v} \langle q=0 | \prod_{I=1}^N
\left( \gamma_I\,
\langle {\cal V}^{bc}_{I}| \gamma^{-1}_I \right) | q=0\rangle_{v}~,
\eeq
where the $N$ Hilbert spaces labeled by $I$ describe the external
fields and are all independent, while $\gamma_I$ are (arbitrary)
projective transformations mapping the interaction point from $z=1$ as
in~\eq{V30} to the arbitrary point $z_I$. Remember that the vertex
in~\eq{tree} contains the tensor product of the $N$ vacua ${}_I
\langle 0; q=-Q|$, but for notational simplicity we write it simply as
$\langle V^{bc}_{N;0}|$. Its explicit form is written in Eq.~(2.26)
of~\cite{DiVecchia:1989id} in terms of the representation $E^\l$ with
weight $\l$ of the projective group. As we have seen in the previous
$1$--loop example, the partition function of a fermionic system is in
general vanishing because of the presence of zero--modes. For higher
genus surfaces, the Riemann--Roch~\eq{R-R} theorem shows that this
happens also for non--trivial periodicity conditions; so the simplest
non--trivial amplitude is the correlation function
\beq\label{g-l}
Z_\e^{\l}(z_1,\ldots,z_{N_b}) = \langle \prod_{I=1}^{N_b} b(z_I)
\rangle_{(\e,\l)} = \langle V^{bc}_{N_b;g} | \prod_{I=1}^{N_b}\left(
b^{(I)}_{-\l} | q=0\rangle_I \right) ~.
\eeq
where $\langle V^{bc}_{N_b;g} |$ is the generalization of the
vertex~\eq{tree} at genus $g$. Thus $Z_\e^{\l}$ is the Green function
on a surface of genus $g$ with non--trivial boundary conditions along
the $b$--cycles and with $N_b = |Q|\,(g-1)$ insertions of the $b$
field. In order to construct $ \langle V^{bc}_{N_b;g} |$ within the
sewing approach, we use the generating vertex $\langle V^{bc}_{N;0}|$
with $N=2 g+ N_b$ and then identify $g$ pairs of Hilbert spaces
(labeled by the index $\mu=1,\ldots, 2 g$) by means of the twisted
propagator that was used also in the torus computation~\eq{perio}.
Moreover we are interested in the ``minimal'' correlation
function~\eq{g-l} that contains only $b$--fields as external states
(labeled by the index $I=1,\ldots,N_b$). So we can set to zero all the
$b^{(I)}$ oscillators in the computation of $\langle V^{bc}_{N_b;g}|$
and keep only the $c^{(I)}$ fields that will be saturated by the $N_b$
$b$--insertions of~\eq{g-l}. Here we report just the final result for
the generating vertex $\langle V^{bc}_{N_b;g}|$, while the details of
the construction are postponed to the Appendix~C
\begin{equation}\label{nC37}
\langle{V^{bc}_{N_b,g}}| = {\rm det}(1-H) \left[\prod_{I=1}^{N_b}
{}_{I}\langle{q=-Q}|\right] {\cal F}_{N_b,g}~.
\end{equation}
The formal structure of this equation is quite natural. The non--zero
mode contribution is given by a (fermionic) Gaussian integral and
this is why the determinant ${\rm det}(1-H)$ appears. As in the usual
case $\e_\mu=0$, also for our computation this determinant can be
written in terms of the multipliers of the Schottky group
\begin{equation}\label{tdetH}
{\rm det}(1-H) = {\prod_{\a}}' \prod_{n=\lambda}^\infty
(1-\ex{-2\pi\ii \e \cdot N_\a} k^n_\alpha)
(1-\ex{2\pi\ii \e \cdot N_\a} k^n_\alpha)~.
\end{equation}
$N_\a$ is a vector with $g$ integer entries; the $\mu^{\rm th}$ entry
counts how many times the Schottky generators $S_\mu$ enters in the
element of the Schottky group $T_\a$, whose multiplier is $k_\a$; the
appearance of each $S_\mu$ is counted with the exponent sign, so that
$S_\mu$ contributes~$1$, while $(S_\mu)^{-1}$ contributes~$-1$ to the
global value of $N_\a^\mu$. The product ${\prod_{\a}'}$ is over the
primary classes of the Schottky group excluding the identity and
counting $T_\a$ and its inverse only once. Eq.~\eq{tdetH} is the
generalization of the first product of~\eq{t1l} for an orientable
Riemann surface of genus $g$.

The second term of~\eq{nC37} takes into account the zero--mode
contribution. In this case it is not possible to write a simple
expression that is valid for all arbitrary $\l$. The complication
arises because the zero--mode $b_s^{(2\mu)}$ are entangled with the
external oscillators $c^{(I)}_\l$. Thus it is not possible to derive
the zero--mode contribution to the $g$--traces before having computed
the scalar product over the $N_b$ external Hilbert spaces present in
\eq{g-l}. This technical problem appears clearly from the formulae of
the Appendix~C. Thus we now consider directly the correlation
functions~\eq{g-l}.
By inserting Eq.~\eq{nC37} into~\eq{g-l}, one gets
\beq\label{g-l:q1}
Z_\e^{\l}(z_1,\ldots,z_{N_b})
\equiv {\rm det}(1-H) \;{\cal F}^{(\l)}
~,
\eeq
where
\beq\label{ncalf}
{\cal F}^{(\l)} =\left( \prod_{I=1}^{N_b} {}_I\langle q=-Q|\right)
{\cal F}_{N_b,g} \left( \prod_{I=1}^{N_b}b^{(I)}_{-1} |q=0 \rangle_I
\right) ~.
\eeq

\subsubsection*{The case $|Q|=1$}
Let us first focus on the case $Q=-1$, where the conformal weight of
$b(z)$ is $\l=1$ and the one of $c(z)$ is zero; we consider generic values
for the twist $\e_\mu$. As explained previously, the non--zero mode
contribution is given by~\eq{tdetH} with $\l=1$. On the other hand, in
the Appendix~C, we show that the contribution of zero--modes contained
in ${\cal F}$ (see~\eq{ncalf}), which depends on $g-1$ variables $z_I$,
is naturally written as a determinant
\beq\label{detq1}
{\cal F}^{(1)} ={\rm det}\left(
\begin{tabular}{ccc}
$\zeta_1(z_1)$ & \ldots & $\zeta_g(z_1)$ \\
\vdots & & \vdots \\
$\zeta_1(z_{g-1})$ & \ldots & $\zeta_g(z_{g-1})$ \\
$\ex{2\pi\ii \e_1} - 1 $  & \ldots &
$\ex{2\pi\ii \e_g} - 1 $
\end{tabular}
\right)~,
\eeq
with
\beq\label{Om}
\zeta_\mu(z_I) = \sum_\a \ex{2\pi\ii (\e\cdot N_\a
+\e_\mu)}\left[\frac{1}{z_I - T_\a S_\mu(0)}-\frac{1}{z_I - T_\a(0)}
\right]~,
\eeq
where the sum runs over the all the elements of the Schottky group.
Thus from Eq.~\eq{nC37} one obtains
\beq\label{re1}
Z_\e^{\l=1}(z_1,\ldots,z_{g-1}) = {\cal F}^{(1)}~ {\prod_{\a}}'
\prod_{n=1}^\infty  (1-\ex{-2\pi\ii \e \cdot N_\a} k^n_\alpha)
(1-\ex{2\pi\ii \e \cdot N_\a} k^n_\alpha)~.
\eeq
The functional form of this result is exactly the expected one: in
fact one can interpret the second term with the Schottky product as
the non--zero mode contribution to the determinant of the Dirac
operator and the first determinant as the zero--mode contribution to
the correlation function $Z_\e^{\l=1}$, written in terms of
twisted abelian differentials. However, even if this interpretation
will be eventually the correct one, it can not be applied to
Eq.~\eq{re1} as it stands. First Eq.~\eq{Om} defines $g$ functions
$\zeta$, while, according to the Riemann--Roch~\eq{R-R} theorem we
have only $g-1$ abelian differentials. Moreover, it is annoying that
the origin of the complex plane plays a privileged r\^ole
in~\eq{Om}. In particular this means that the $\zeta$'s just
introduced are not regular: in fact, when $T_\a$ is the identity or
$S_\m^{-1}$, they have a pole for $z_I=0$, which in general is part of
the fundamental domain representing the Riemann surface. It turns out
that one can solve these problems simultaneously.

If all the twists $\e_\mu$ are trivial, then the determinant is zero
because of the last line. This is of course due to the extra
zero--mode ($c(z)\Leftrightarrow const$) appearing in the untwisted case
which makes the
correlator~\eq{g-l:q1} vanish. So let us suppose that at least one
$\e$ is non--trivial, for instance $\e_g\not= 0$. With this hypothesis
it is easy to simplify~\eq{detq1} by making a linear combination of
each column with the last one so to set to zero the first $g-1$
entries of the last row
\beq\label{detq2}
{\cal F}^{(1)} = {\rm det}\left(
\begin{tabular}{cccc}
$\Omega_1(z_1)$ & \ldots & $\Omega_{g-1}(z_1)$ & $\zeta_g(z_1)$ \\
\vdots & & \vdots & \vdots\\
$\Omega_1(z_{g-1})$ & \ldots & $\Omega_{g-1}(z_{g-1})$ &
$\zeta_g(z_{g-1})$ \\
$0$  & \ldots & $0$ &$\ex{2\pi\ii \e_g} -1 $
\end{tabular}
\right)~,
\eeq
where
\beq\label{tda}
\Omega_\mu (z_I) = 
\zeta_\mu (z_I) - \frac{\ex{2\pi\ii \e_\mu}-1}{\ex{2\pi\ii \e_g}-1}\;
\zeta_g (z_I) 
~~,\quad \mu=1,\ldots,g-1~. 
\eeq
This shows that only the $g-1$ functions $\Omega$ enter in the final
result~\eq{re1} so that it is natural to identify these $\Omega$'s
with the twisted abelian differentials. Let us shows that they have
all the expected properties. First, the dependence on the origin
disappeared and, contrary to the original $\zeta$'s, the $\Omega$'s
are everywhere regular. To see this, it is useful to rewrite the
sum~\eq{Om} in a different form\footnote{See the Appendix~C for the
derivation.}, separating the contributions coming from the Schottky
elements of the form $T_\a S_\mu^l$
\bea\label{si1}
\zeta_\mu(z_I) & = & {\sum_\a}^{(\mu)} \ex{2\pi\ii (\e\cdot N_\a
+\e_\mu)} \left[\frac{1}{z_I - T_\a(\eta_\mu)} -
\frac{1}{z_I - T_\a(\xi_\mu)} \right]
\\ \nonumber & + &
 (1-\ex{2\pi\ii\e_\mu}) \sum_\a \ex{2\pi\ii \e\cdot N_\a}
\left[\frac{1}{z_I - T_\a(0)} - \frac{1}{z_I - T_\a(a_\mu^\a)}
\right]~,
\eea
where, in the second line, $a_\mu^\a = \eta_\mu$ if the $T_\a$ is of
the form $T_\a= T_\beta S_\mu^l$ with $l\geq 1$, while $a_\mu^\a =
\xi_\mu$ otherwise. It is also important to remember that the sum in
the first line does not contain all the Schottky elements whose
rightmost generator is $S_\mu^{\pm 1}$, while the second sum is over
all the elements $T_\a$. It is now easy to see that first term in the
second square bracket of~\eq{si1} cancels in the
combination~\eq{tda}. Then one can check the periodicity
properties. It is clear that all the expressions we have written so
far are periodic along the $a$--cycles ($z -\eta) \to \ex{2\pi\ii}
(z-\eta)$, since they are holomorphic in~$z$. The periodicity along
the $b$--cycles can be deduced by looking at the Eq.~\eq{Om}. By means
of the identity~\eq{ideu} one can see that
\beq\label{Bcp}
\zeta_\mu(S_\nu(z))\;dS_\nu(z) = \ex{2\pi\ii \e_\nu} \zeta_\mu(z)\;dz~.
\eeq
Since the $\Omega$'s are simply linear combinations of the $\zeta$'s,
they have the same periodicity property, which is exactly what one
expected from Eq.~\eq{perio}. The fermionic result~\eq{re1} is
multilinear in the abelian differentials $\Omega$'s and thus has a
non--trivial periodicity along the $b$--cycles~\eq{Bcp}
\beq\label{fpe}
Z_{\e}^{\l=1}(z_1,\ldots,S_\nu(z_k),\ldots,z_{g-1})\;S'_\nu(z_k) =
\ex{2\pi\ii \e_\nu}\, Z_{\e}^{\l=1}(z_1,\ldots,z_{g-1})~.
\eeq

Notice that, if some $\e_\mu = 0$ (of course with $\mu\not= g$), the
corresponding $\Omega_\mu$ reduces to $\zeta_\mu$, where the second
line of~\eq{si1} vanishes. This is the na\"\i ve generalization of the
usual untwisted abelian differentials $\omega_\mu(z_I)$ (given in~\eq{uab})
  with an
additional phase $\exp{(2\pi\ii \e\cdot N_\a})$, which ensures the
twisting of the periodicity along the cycles $b_\nu$ with
$\nu\not=\mu$. For the general case $\e_\mu\not=0$, the twisted
periodicity along $b_\mu$ requires also the second line of
Eq.~\eq{si1}.

\subsubsection*{The case $|Q| \geq 2$}
The generalization of the result just presented to the case
$\l=3/2,2,\ldots$ is discussed in Appendix~C. We recall here the main
qualitative features. ${\cal F}^{(\l)}$~\eq{detq1l} has the same
structure of ${\cal F}^{(1)}$, but the elements of this new matrix are
themselves $(2\l-1)\times (2\l-1)$ blocks. This is simply because the
zero--modes now are related to the oscillators $b_r$, with
$r\in[(1-\l),(\l-1)]$. The presence of many zero--modes implies that,
in the ``minimal'' correlation function~\eq{g-l} there are $(2\l-1)
(g-1)$ variables $z_I$ and that the building blocks of the matrix
${\cal F}^{(\l)}$ have extra indices $r,s \in[(1-\l),(\l-1)]$ and
become $\zeta_{\mu,s}(z_{\nu,r})$. Also the entries of the last row in
${\cal F}^{(\l)}$ are now replaced by $(2\l-1)\times (2\l-1)$ matrices
\beq\label{cale}
{\cal E}(S_\m)_{rs} = \left(\ex{2\pi\ii \e_\m} E(S_\mu)
-\!\!\one \right)_{rs} ~.
\eeq
The $\zeta_{\mu,s}(z_{\nu,r})$'s suffer the same diseases discussed in
the $\l=1$ case. These problems are cured in a similar way by defining
$(2\l-1) (g-1)$ $\l$--differentials $\Omega_{\m,s}(z_{\nu,r})$,
analogous to the ones introduced in~\eq{tda} (see Eq.~\eq{tdal}). An
important difference with the respect to the $\l=1$ case is that, for
$\l>1$ and $g>1$, the determinant ${\cal F}^{(\l)}$ does not vanish
even if $\e_\m=0$, $\forall \m$. Therefore the correlation
function~\eq{g-l:q1} is non--trivial also when all the $\e_\m$ are set
to zero. Notice that, if $\l$ is half-integer, then the
expression~\eq{tdal} can be used also in absence of twists
($\e=0$). In fact in this case the matrix ${\cal E}(S_\m)_{rs}$ is
always invertible. For integer $\l$, on the contrary, the determinant
of ${\cal E}(S_\m)_{rs}$ is vanishing for $\e=0$ and the form of the
$\l$--differentials is more complicated, as discussed
in~\cite{DiVecchia:1989id}. Of course one can still express all the
correlation function in terms of the $\zeta_{\mu,s}(z_{\nu,r})$
introduced in~\eq{zgl}.

\sect{Comparison with bosonic determinants}

We are now in the position to generalize Eq.~\eq{2.17}. As anticipated
in the Introduction, this kind of relations can be viewed as a
consequence of the equivalence between fermionic and bosonic
systems. In fact, using again the sewing technique, it is possible to
derive the bosonic correlation functions corresponding to
$Z_\e^{\l}$ of Eq.~\eq{g-l}. In the notation
of~\cite{DiVecchia:1988cy} these correlators are
\beq\label{bg-l}
Z_{\e_b}^{Q}(z_1,\ldots,z_{N_b}) = \langle \prod_{I=1}^{N_b}
:\ex{-\phi (z_I)}: \rangle_{(\e_b,Q)} = \langle V^{\phi}_{N_b;g}
| \prod_{I=1}^{N_b}\Big( 
| q=-1\rangle_I \Big) ~,
\eeq
where the bosonic system has background charge $Q=1-2\lambda$.
Moreover the twisted boundary conditions along the $b$--cycles are
enforced on the bosonic side by the insertion together with the
propagator of the operator $\ex{2\pi\ii \e_b
p_0}$~\cite{DiVecchia:1988cy}. Notice, however, that $\e_b = \e -
1/2$, where the additional factor of $1/2$ is necessary in order to
reproduce in the bosonic language the usual $(-1)^F$--twist of the
fermionic traces (see for instance Appendix~A of~\cite{Polchinski:rq},
Eq.~(A.2.23)).

In order to provide a simple example, let us write the $1$--loop
partition function of a bosonic system with $Q=1-2\l$ with twist
$\e_b$ (see, for instance, Eq.~(3.21) of~\cite{DiVecchia:1988cy} with
$\alpha=0$, $\beta=\e_b$ and $N_1=N_2=0$)
\beq\label{t1liX}
Z_{\e_b} = \prod_{n=1}^\infty \left(1-k^n\right)^{-1}\;
\theta\left(\e_b - Q \left(\frac{1}{2} + \Delta\right)
\Big|\;\tau\right)~.
\eeq
This formula has to be equal to Eq.~\eq{t1l} for $\e_b = \e - 1/2$ and
from this relation one can derive the usual product formula for the
$1$--loop Theta function. In particular, in terms of the odd
$\theta_1$, one has
\beq\label{t11-l}
\theta_1(\e|\tau) = 2 k^{1/8} \sin{\pi\e} \prod_{n=1}^\infty
\left(1-k^n\right)
\left(1 - \ex{2\pi\ii \e} k^{n} \right)
\left(1 - \ex{-2\pi\ii \e} k^{n} \right)~.
\eeq
The higher genus case works in a similar way. The main difference is
that the partition function without insertions now vanishes. On the
fermionic side we know that this is due to the presence of zero-modes,
while on the bosonic side this is a simple consequence of the momentum
conservation modified by the background charge. Thus, in order to
obtain a non--trivial result, we have to compare the correlation
functions~\eq{g-l:q1} and \eq{bg-l}. For $Q=-1$ the fermionic result
has been obtained in Section~4 and the bosonic one\footnote{In
Appendix~A we recall the definition and the main properties of the
various geometrical objects present in~\eq{reb1}.} is (again from
Eq.~(3.21) of~\cite{DiVecchia:1988cy})
\beq\label{reb1}
Z_{\e_b}^{Q=-1} (z_1,\ldots,z_{g-1}) =
\frac{\prod_{I=1}^{g-1} \sigma (z_I) \prod_{I<J} E(z_I,z_J)}
{{\prod_{\a}'} \prod_{n=1}^\infty (1- k^n_\alpha)}
\;\theta\!\left(\e +\Delta - \sum_{I=1}^{g-1} J(z_I)
\,\Bigg|\, \tau \right)\;,
\eeq
where we have used again $\e_{b\,\mu} = \e_\mu -1/2$ $\forall
\mu=1,\ldots,g$. Notice that, thanks to~\eq{Jm} and~\eq{deltaz0}, the
argument of the Theta--function on the r.h.s. does not depend on the
basis point $z_0$ of the Jacobi map.
The explicit expressions in terms of the Schottky parametrization of
$\sigma(z_I)$ (including its normalization), $E(z_I,z_J)$ and $\Delta$
are given in Appendix~A of~\cite{DiVecchia:1988cy}.

The equivalence with the fermionic result
implies\footnote{From now on, we endow the $\l$--differentials derived
in the sewing approach with the appropriate factors of $dz^\l$.}
\eq{re1}$\;=\;$\eq{reb1}
\beq\label{t2.17}
C_\e^{(1)}~ {\cal F}^{(1)} =
\prod_{I=1}^{g-1} \sigma (z_I) \prod_{I<J} E(z_I,z_J)
\;\theta\!\left(\e + \Delta  - \sum_{I=1}^{g-1} J(z_I)
\,\Bigg|\, \tau \right)\;,
\eeq
which generalizes~\eq{2.17} to the twisted case, with
\beq\label{Ce}
C_\e^{(1)} = {\prod_{\a}}' \prod_{n=1}^\infty (1- k^n_\alpha)
(1-\ex{-2\pi\ii \e \cdot N_\a} k^n_\alpha)
(1-\ex{2\pi\ii \e \cdot N_\a} k^n_\alpha)~.
\eeq

A first check of this identity comes from the study of the periodicity
properties of the results. The periodicity~\eq{fpe} is reproduced in
the bosonic side in a complicated way by combining the various factors
coming from transformation of the objects on the
r.h.s. of~\eq{reb1}. However most of these contributions cancel as in
the untwisted case and the only $\e$-dependent contribution comes
from the Theta function. From Eqs.~\eq{thetaper}, \eq{pmbs}
and~\eq{sigbs}, one gets
\beq\label{bpe}
Z_{\e_b}^{Q=-1}(z_1,\ldots,S_\nu(z_k),\ldots,z_{g-1})\; S'_\nu(z_k) =
\ex{2\pi\ii \e_\nu}\, Z_{\e_b}^{Q=-1}(z_1,\ldots,z_{g-1})~,
\eeq
in agreement with~\eq{fpe}. Actually also a stronger check
of~\eq{t2.17} is possible. Since we know the explicit form for the
twisted abelian differentials~\eq{tda}, we can use it together with
the expressions contained in Appendix~A of~\cite{DiVecchia:1988cy} in
order to express both sides of~\eq{t2.17} in terms of the Schottky
group. At this point one can develop the results for small $k_\mu$ and
keep only a finite number of terms in all the series or infinite
products so to obtain in both sides a polynomial in the $k_\mu$'s. The
coefficients of each term are (complicated) functions of the fixed
points and of the twisting parameters $\e$'s. A highly non--trivial
check is to verify that the coefficients obtained on the
l.h.s. of~\eq{t2.17} agree with those obtained from the expansion of
the r.h.s. We performed this check for the case $g=2$ and verified
that the first four terms in the expansion of~\eq{t2.17} actually
agree.

The generalization of~\eq{t2.17} to the case $|Q|\geq 2$ is
straightforward and reads\footnote{With a small abuse of notation,
the term of $1/2$ in~\eq{t2.17Q} stays for a $g$--component vector
whose entries are all $1/2$.}
\beq\label{t2.17Q}
 C_\e^{(\l)}~ {\cal F}^{(\l)}=
\frac{ \prod_{I<J} E(z_I,z_J) }{\prod_{I=1}^{N_b} \sigma (z_I)^Q}
\;\theta\!\left(\e -\frac{1}{2} - Q\left( \Delta + \frac{1}{2}\right)
- \sum_{I=1}^{N_b} J(z_I)
\,\Bigg|\, \tau \right)\;,
 \eeq
with
\beq\label{CeQ}
C_\e^{(\l)} = {\prod_{\a}}'\left( \prod_{n=1}^\infty (1- k^n_\alpha)
 \prod_{n=\l}^\infty(1-\ex{-2\pi\ii \e \cdot N_\a} k^n_\alpha)
(1-\ex{2\pi\ii \e \cdot N_\a} k^n_\alpha)\right)~.
\eeq

\sect{Conclusions}

In this paper we exploited the sewing technique to study free fermions
of arbitrary spin (and free boson with arbitrary background charge)
with twisted boundary conditions on the $b$--cycles. We computed
various quantities of interest like determinants and twisted
differentials and gave explicit expressions for them in terms of the
Schottky uniformization of the Riemann surface. By comparing the
bosonic and the fermionic results for certain ``minimal'' correlation
functions we also found new identities between products of multipliers
over the Schottky group and the usual multiloop Theta functions.
Identities of this kind are important because they make manifest the
properties under modular transformations which are hidden when the
results are written in Schottky parametrization. This is very useful,
for instance, in string theory, where the modular properties of
perturbative amplitudes are crucial for the consistency of the
theory. On the other hand, important building blocks of string
amplitudes, like the partition functions, are often given in terms of
Schottky product like~\eq{Ce} and thus their modular properties are
blurred. It is clearly important to rewrite these products in terms of
Theta--functions.  Let us mention here an example of this problem in
the context of open bosonic strings~\cite{wip}.

It is well known that a system of $g+1$ D-branes, connected among them
with tubes representing the closed string propagation, is equivalently
described by a disk with $g$ holes. The two descriptions make manifest
different unitarity properties of the result: in the D-brane picture
the poles of the amplitude in the Schottky multipliers are related to
the propagation of some almost on--shell {\em closed string} states
between  D-branes, while in the disk description the poles are
related to {\em open string} states. The difference is made more
evident if one switches on a constant gauge field strength $F$ on the
D-branes (or, in the $T$--dual language, if some of the D-branes in
this multi--body system have a non--vanishing constant relative
velocity). In the D-brane language the result can be derived by
introducing in the construction of~\cite{Frau:1997mq} a twist similar
to the one discussed in this paper. In this case $\e$ is related to
the differences between the external fields $F$. In the D-brane
description these twists enter only as a phase independent of the
moduli, exactly as in~\eq{Ce}. By using the results of this paper one
can explicitly perform the modular transformation of this expression
and obtain the same quantity in the open string description, where the
surface looks like a disk with $g$ holes. Now the phases induced by
the twists do depend on the moduli and this is strictly related to the
modification induced by $F$ on the mass of the open strings. This very
important difference is already present at $1$--loop level ({\em i.e.}
in a system of two D-branes), as it is clear by the comparison between
the results of~\cite{Bachas:bh} and~\cite{Billo:1997eg}. We have
checked~\cite{wip} that, as already suggested in~\cite{Bachas:bh},
from the open string formulation of the charged partition function one
can derive the $1$--loop Euler--Heisenberg effective action for
Yang--Mills theories simply by performing the field theory limit
 $\a' \to 0$  of the results, in the spirit of~\cite{Frizzo:2000ez} (see
also references therein). Of course one obtains the pure Yang--Mills
effective action if the bosonic D3-branes are used as starting point,
or the ${\cal N}=4$ result if one starts from the usual type IIB
D3-branes. This technique can be extended to more complicated systems
of ${\cal N}=2$ theories with various content of matter~\cite{mb},
once the corresponding string results are derived. On the contrary,
the description of~\cite{Billo:1997eg} and its bosonic multiloop
version~\cite{Frau:1997mq} are not directly connected to the
Yang--Mills effective action. In this case the low energy limit
captures the (super)gravity interaction among the D-branes seen as
classical gravitational solitons.
Notable exceptions to this are known, where {\em perturbative} gauge
theory results are directly connected to supergravity
computations. This striking connection is usually due to the
cancellation of the contribution related to the stringy modes. In
fact, if the complete string result is due only to massless modes
exchange (of the closed and the open strings respectively in the two
descriptions), then the usual open/closed string duality connects
directly gauge theory quantity to supergravity
tree--diagrams~\cite{DiVecchia:2003ae}. Of course this does not happen
for the whole effective action even in the maximally supersymmetric
models, but it is a possibile feature of only the first non--vanishing
term in the small $F$ (or small $\e$) expansion.

Coming back to the technical construction presented in this paper, the
main open problem is to extend the sewing approach to the case where
one has twists along the $a$--cycles. As just argued, one can avoid
this problem for certain cases by exploiting the modular properties of
the results. However, modular transformations cannot be enough when
twists are present both along the $a$--cycles and the $b$--cycles at
the same time. This seems a challenging problem, since even in the
simplest situations, one looses the usual relation between string
amplitudes and the Schottky group. The main difficulty is that the
building blocks of the amplitudes have not been expressed so far in
terms of representation of the projective group,
see~\cite{Chu:2002nd}.

\section*{Acknowledgments}

We would like to thank C.S. Chu for collaboration on related topics
and for interesting discussions and suggestions on the results
presented here. We would like to thank Andrew McIntyre for email
exchange and for pointing out to us various interesting papers in the
mathematical literature.  This work is supported in part by EU RTN
contract HPRN-CT-2000-00131 and by MIUR contract 2001-025249.  R.R. is
supported by European Commission Marie Curie Postdoctoral Fellowships.

\section*{Appendix A: Definitions}
\renewcommand{\theequation}{A.\arabic{equation}}
\setcounter{equation}{0}

In this Appendix, we collect the definitions of some quantities of
relevant interest in the theory of Riemann surfaces.

Let $\Sigma$ be a compact Riemann surface of genus $g$ with a complex
structure defined on it. There exists a normalized basis of
holomorphic $1$--forms $\omega_\mu$ ($\mu=1,\ldots,g$), called
\underline{\em abelian differentials}, such that
\beq\label{nda}
\frac{1}{2\pi \ii} \oint_{a_\mu} \omega_\nu = \delta_{\mu\nu}~~,~~~
\frac{1}{2\pi \ii} \oint_{b_\mu} \omega_\nu = \tau_{\mu\nu}~,
\eeq
where $a_\mu$ and $b_\mu$ are a canonical basis of homology cycles
with intersection matrix: $I(a_\mu,b_\nu)=\delta_{\mu\nu}$,
$I(a_\mu,a_\nu)=I(b_\mu,b_\nu)=0$. The matrix $\tau$ is called
\underline{\em period matrix} on $\Sigma$ and is a symmetric matrix
with positive definite imaginary part.

Given a base point $z_0 \in \Sigma$, one can associate to any $z \in
\Sigma$ a complex $g$--component vector $\vet{J}(z)$ by means of the
\underline{\em Jacobi map}
\beq \label{Jm}
\vet{J}:~z \to J_\mu(z)= \frac{1}{2\pi \ii} \int_{z_0}^z \omega_\mu~.
\eeq
The vector $\vet{J}$ is defined up to periods around $a_\mu$ or
$b_\mu$~\eq{nda} and so belongs to the complex torus $J(\Sigma) =
{\mathbb C}^g/({\mathbb Z}^g + \tau {\mathbb Z}^g)$ called
\underline{\em Jacobi variety}.

The \underline{\em Riemann Theta function} associated to the surface
$\Sigma$ is defined for $\vet{Z}\in {\mathbb C}^g$ by
\beq\label{thetafun}
\theta(\vet{Z}|\tau) = \sum_{n\in {\mathbb Z}^g} \exp\{\ii \pi
\vet{n}\cdot \tau \cdot \vet{n} +2 \pi \ii \vet{n} \cdot \vet{Z} \}~,
\eeq
where $\vet{n}\cdot \tau \cdot \vet{n} = \sum_{\mu,\nu=1}^g n_\mu
\tau_{\mu\nu} n_\nu$ and $\vet{n} \cdot \vet{Z} = \sum_{\mu=1}^g n_\mu
Z_\mu$. The Theta function has a simple transformation law under
shifts of $Z$ on the lattice ${\mathbb Z}^g + \tau {\mathbb Z}^g$:
\beq\label{thetaper}
\theta(\vet{Z}+\tau\cdot \vet{n} + \vet{m}|\tau) = \exp[ -\pi\ii
\,\vet{n} \cdot \tau \cdot \vet{n} - 2 \pi \ii \vet{n}\cdot \vet{Z}]\;
\theta(\vet{Z}|\tau)~.
\eeq

The \underline{\em Riemann vanishing theorem} states that
$\theta(\vet{Z}|\tau)$ vanishes if and only if there exist $g-1$
points ($z_1,\ldots,z_{g-1}$) on $\Sigma$ such that
\beq\label{rvt}
\vet{Z} =  \vet{\Delta}_{(z_0)} - \sum_{\rho=1}^{g-1}
\vet{J}(z_\rho)~,
\eeq
where $\vet{\Delta}_{(z_0)}$ is a constant vector in $J(\Sigma)$
(called \underline{\em Riemann class}) and the vectors $J(z_\rho)$ are
defined by~\eq{Jm}. The dependence of $\Delta_{(z_0)}$ from the base
point $z_0$ is given by:
\beq\label{deltaz0}
\vet{\Delta}_{(z)} = \vet{\Delta}_{(z_0)} - \frac{g-1}{2\pi\ii}
\int_{z_0}^z \vet{\omega}~.
\eeq
We also used two other important functions defined on $\Sigma$:
$E(z,y)$ and $\sigma(z)$.

The \underline{\em prime form $E(z,y)$} is completely defined by the
following properties:
\begin{itemize}
\item it is a holomorphic differential form on $\Sigma \otimes \Sigma$
of weight $(-1/2,0) \times (-1/2,0)$;
\item it is odd under the exchange $z \leftrightarrow w$;
\item it has a simple zero for $z \to w $ with the normalization
\beq \label{ezw}
E(z,w) \underset{z\to w}\longrightarrow
\frac{z-w}{\sqrt{d z}\sqrt{d w}}~;
\eeq
\item it is single valued around the $a$--cycles, but not around the
$b$--cycles: if $z$ goes around $b_\mu$ the prime form is shifted in
the following way
\beq \label{pmbs}
E(z,w) \longrightarrow -\exp[-\pi\ii \tau_{\mu\mu} -\int_w^z
\omega_\mu]~ E(z,w)~.
\eeq
\end{itemize}
The weight $(-1/2,0) \times (-1/2,0)$ means that $E(z,w)$ contains the
square root of the differentials $dz$, $dw$ at the denominator. The
r\^ole of these differentials becomes clear on the sphere, where
$E(z,w)$ is exactly given by the r.h.s. of Eq.~\eq{ezw}. In fact the
differentials make $E(z,w)$ invariant under inversion: $z\to 1/z$,
$w\to 1/w$ and therefore assure a regular behaviour of the prime form
at the infinity.

Finally \underline{\em the function $\sigma(z)$} is defined up to a
constant factor by the following properties
\begin{itemize}
\item it is a holomorphic differential form on $\Sigma$ of weight $g/2$,
\item it has no zeros,
\item it is single valued around the $a$--cycles and transforms as
\beq\label{sigbs}
\sigma(z) \longrightarrow \exp[\,\pi\ii(g-1) \tau_{\mu\mu} +\pi\ii-2\pi\ii
(\Delta_{z})_\mu ]\; \sigma(z)~,
\eeq
when $z$ is moved around the cycle $b_\mu$.
\end{itemize}

In the second part of this Appendix we will discuss the properties of
these quantities under the modular transformations, that is under a
change of the canonical homology basis. Two homology basis $(a,b)$ and
$(\tilde{a},\tilde{b})$ are related by a modular transformation if
there is a $2g\times 2g$ matrix with integer entries
\beq \label{sp2gz}
\left(
\begin{tabular}{cc}
$A$ & $B$ \\ $C$ & $D$
\end{tabular}
\right) \in Sp(2g, {\mathbb Z})~,
\eeq
such that
\beq \label{taa}
\left(
\begin{tabular}{c}
$\tilde{b}$ \\ $\tilde{a}$
\end{tabular}
\right) =  \left(
\begin{tabular}{cc}
$A$ & $B$ \\ $C$ & $D$
\end{tabular}
\right)
\left(
\begin{tabular}{c}
$b$ \\ $a$
\end{tabular}
\right)~.  \eeq The $g\times g$ blocks of the matrix in~\eq{sp2gz}
satisfy the constraints $AB^t=BA^t$, $CD^t=DC^t$, $AD^t-BC^t=1_g$,
which assure the invariance of the intersection matrix\footnote{Notice
  that the square of the intersection matrix is $-1$, which implies
  that also the transpose of the matrix in~\eq{sp2gz} belongs to
  $Sp(2g, {\mathbb Z})$; so we also have $A^t\,C=C^t\,A$,
  $B^t\,D=D^t\,B$, $A^t\,D-C^t\,B=1_g$.}. In order to preserve the
canonical normalization of the abelian differentials~\eq{nda}, the
$\tilde{\omega}$ in the new basis are related to the old ones by
\beq \label{mtad}
\tilde{\omega} = \omega\cdot (C \tau + D)^{-1}~,
\eeq
and correspondingly one gets the new period matrix $\tilde{\tau}$
\beq \label{mtpm}
\tilde{\tau}_{\mu\nu} = \frac{1}{2\pi\ii} \oint_{\tilde{b}_\mu}
\tilde{\omega}_\nu =[ (A \tau + B)(C \tau +D)^{-1}~]_{\mu\nu}.
\eeq
The transformation of the prime form under \eq{taa} is given, for
example, in~\cite{fay2}
\beq\label{mtpf}
\tilde{E}(z,w) = \exp\left[ \frac{1}{4 \ii\pi}
\int_z^w \omega \cdot (C\tau+D)^{-1}
C \cdot\int_z^w \omega \right] E(z,w)~,
\eeq
The transformation properties of the Riemann class~\eq{deltaz0}, the
Riemann Theta function~\eq{thetafun} and the functions $\sigma$ depend
on the values of the diagonal elements of $C D^t$ and $A B^t$. Here we
focus on the case where $(CD^t)_{\m\m}=(A B^t)_{\m\m} = 2{\mathbb Z}$ ,
 $\forall \m=1,\ldots,g$ . The Riemann
class~\eq{deltaz0} has a simple transformation property under these
elements of the modular group
\beq\label{mtjd}
\tilde{\Delta} = \Delta (C\tau + D)^{-1}~.
\eeq
The Riemann Theta function~\eq{thetafun} transforms as
\beq\label{mdtf}
\theta(\tilde{Z}|\tilde{\tau}) = {\xi}\,{\sqrt{{\rm det} (C\tau +
D)}} \; \ex{\ii\pi Z\cdot (C\tau+D)^{-1} C \cdot Z} \theta(Z|\tau)~,
\eeq
where $\tilde{Z} = Z (C\tau + D)^{-1}$ and $\xi$ is a phase such
that $\xi^8=1$ depending only on the modular transformation.
Finally the function $\sigma$ undergoes the following transformation:
\beq\label{mtsi}
\tilde{\sigma}(z) = K\, \exp\left[- \frac{\ii\pi}{g-1} \Delta_z \cdot
(C\tau+D)^{-1} C \cdot \Delta_z \right] \sigma(z)~,
\eeq
where $K$ is a normalization factor independent of $z$.

\section*{Appendix B: The Schottky parametrization}
\renewcommand{\theequation}{B.\arabic{equation}}
\setcounter{equation}{0}

It is well known that one can represent the sphere through the
stereographic projection as the extended complex plane:
${\mathbb C} \cup  \, \{\,\infty\, \}$.
This equivalence is the starting point to represent any Riemann
surface as part of the extended complex plane by means of the
so--called Schottky uniformization. Here we will focus on closed
orientable surfaces and will give a concrete realization of the
intuitive idea that one can generate higher genus surfaces just by
adding handles to the sphere. As an example let us derive the Schottky
parametrization of the torus. It is sufficient to perform two
operations. First one has to choose in $\mathbb C$ two points $J=-d/c$
and $J'=a/c$ and cut around them two non--overlapping disks $\cal C$
and $\cal C'$ of the same radius $R=1/|c|$. Then one has to identify
the two circles, which are the borders of the disks, so to obtain
again a closed surface (actually this identification can be done with
a twist of an arbitrary angle $\theta$ related to the phase of $c$).
These two operations can be summarized mathematically by introducing
the loxodromic projective transformation $S \in SL(2,{\mathbb C})$
\beq\label{schel}
S = \left(
\begin{tabular}{cc}
$a$ & $b$ \\ $c$ & $d$
\end{tabular}
\right)~~,~~\mbox{ with }~ ad-bc = 1
~~\mbox{ and }~  ({\rm Tr}\,S)^2 \not\in [0,4]
~.
\eeq
This defining transformation is called \underline{\em generator},
while the complete Schottky group $\cal S$ is obtained by all the
possible products of the $S$'s (just $S^n$ with $n\in {\mathbb Z}$ in
the case of the torus). The isometric circles are described by the
equations:
\beq\label{isoc}
\left|\frac{d S}{d z} \right|^{-1/2} \!\!\! =|c z + d| =1 ~~,~~~
\left|\frac{d S^{-1}}{d z} \right|^{-1/2} \!\!\! =|c z - a| =1 ~,
\eeq
in agreement with what we have said above. The requirement of having
two non--overlapping disks ensures that the projective map $S$
introduced in~\eq{schel} is loxodromic. This kind of transformation
can be equivalently characterized by a couple of fixed points
$(\eta,\xi)$
\beq\label{fp}
\lim_{n\to\infty} S^n(z) = \eta~~,~~~
\lim_{n\to\infty} S^{-n}(z) = \xi~~,~~~\forall z \not= \xi,\eta~
\eeq
and by a multiplier $k$
\beq\label{mul}
k = \frac{S(z) - \eta }{S(z) -\xi}\;\frac{z-\xi}{z-\eta} ~~,~~~
{\rm with}~ |k| < 1~,
\eeq
valid $\forall z \not= \xi,\eta$. A generic loxodromic transformation
$S$ can be written in terms of its fixed points and multiplier in a
simple way
\beq
S = \frac{1}{\sqrt{k}\, (\xi-\eta)}\left(
\begin{tabular}{cc}
$\eta-k \xi$ & $-\xi\eta\, (1-k)$ \\ $1-k$ & $k\eta -\xi$
\end{tabular}
\right)~.
\eeq
Notice that $S$ maps any point {\em outside} the circle $\cal C$ into
a point {\em inside} the circle $\cal C'$. On the contrary $S^{-1}$
maps any point {\em outside} the circle $\cal C'$ into a point {\em
  inside} the circle $\cal C$. Thus the fundamental region
representing the torus is simply the extended complex plane minus the
fixed points, modulo the equivalence relation induced by the various
elements of $\cal S$: $({\mathbb C}\, \cup\, \{\infty\} - \{\xi,\eta
\})/{\cal S}$.

In general, one can apply these same ideas to build a surface of genus
$g$: it is sufficient to cut in the extended complex plane $g$ pairs
of {\em non--overlapping} disks and identify the pairs of the
corresponding circles. Clearly this corresponds to the insertion of
$g$ handles on the sphere. Mathematically this construction is
described by $g$ loxodromic projective transformations $S_\mu$ that
generate freely the Schottky group ${\cal S}_g$. As described in the
Appendix~C, in the sewing procedure the generators of the group
naturally arise from combinations of the local coordinates and of the
propagator (see in particular Eq.~~\eq{gen}). So a Riemann surface
with $g$ holes is represented by the extended complex plane, minus all
the fixed points, modded out by the equivalence relation induced by
${\cal S}_g$. In this uniformization of the surface the $a$--cycles
correspond to the circle $\cal C'$ anti--clockwise oriented, while the
$b$--cycles are represented by lines connecting $z$ and $S_\m(z)$
Notice that the
condition of having non--overlapping circles ensures that each element
$T_\a$ of ${\cal S}_g$ different from the identity is a loxodromic
map and so can be characterized by the fixed points $(\eta_\a,
\xi_\a)$ and by the multiplier $k_\a$. In the text we also make use of
the following standard nomenclature:
\begin{itemize}
\item A primitive element in the group is a transformation which can
  {\em not} be written as an integer power of any other element;
\item A conjugacy class is the set of the elements that can be related
  to each other by a cyclic permutation of their constituent factors
  (for instance, $S_1 S_2$ and $S_2 S_1$ belong to the same conjugacy
  class);
\item A primary class is a conjugacy class containing only primitive
  elements.
\end{itemize}

All the quantities living on Riemann surfaces introduced in an
axiomatic way in thep revious appendix can be explicitly written in
terms of Poincar\'e series on the elements of the Schottky group (see
Appendix~A of~\cite{DiVecchia:1988cy}). We report here only the
abelian differentials for a surface of genus $g$
\beq\label{uab}
\omega_\m (z) = {\sum_\a}^{(\m)} \left(
\frac{1}{z-T_\a(\eta_\m)} - \frac{1}{z-T_\a(\xi_\m)}
\right) ~dz~~,~~~\m=1,\ldots,g~.
\eeq
where ${\sum_\a}^{(\m)}$ means that the sum is over all the elements
of the Schottky group that do not have $S_\m^n$, $n\in {\mathbb Z} -
\{ 0\}$, as their rightmost factor and $\eta_\m$, $\xi_\m$ are the
fixed points of the generator $S_\m$.

Finally it is very useful to notice the following identity
\beq\label{crossr}
\frac{T(a) - T(b)}{T(a) - T(d)}~\frac{T(c) - T(d)}{T(c) - T(b)} =
\frac{a- b}{a-d}~\frac{c - d}{c-b}
\eeq
valid for any projective transformation $T$ and for any points $a$,
$b$, $c$ and $d$. This means that the cross ration of \eq{crossr} is
invariant under projective transformation. For instance, this identity
can be used to rewrite the combinations that typically appear in the
expressions for the abelian differentials
\begin{eqnarray}\label{ideu}
T'(z) \left(\frac{1}{T(z)-x} - \frac{1}{T(z)-y}
\right) & = &
\frac{d}{d z} \log\left(\frac{T(z)-x}{T(z)-y}\;\frac{T(z_0)-y}{T(z_0)-x}
\right) \\
&=&\frac{1}{z-T^{-1}(x)} - \frac{1}{z-T^{-1}(y)} ~.
\nonumber
\end{eqnarray}
For $T=S_\nu$, $x=T_\a(\eta_\m)$ and $y=T_\a(\xi_\m)$, Eq.~\eq{ideu}
shows that $\omega_\m$ is periodic when $z$ goes around a cycle $b_\nu$
({\em i.e.} $z\to S_\nu(z)$).

\section*{Appendix C: The twisted sewing}
\renewcommand{\theequation}{C.\arabic{equation}}
\setcounter{equation}{0}

For trivial boundary conditions $\e_\mu =0$, the derivation of the
generating vertex $\langle V^{bc}_{N_b;g}|$ was discussed in great
detail in the Appendix~C of~\cite{DiVecchia:1989id}. So, here we
will refer to those equations (indicated as ($\cal C$.\#) in the
following) and just point out the novelties or differences introduced
by the presence of the twist $\ex{2\pi\ii \e_\m j_0^\m}$ in the
propagators.

The starting point is the tree--level vertex~\eq{tree} with $N=2g+N_b$
Hilbert spaces.
We label the external Hilbert spaces with the index $I=1,\ldots,N_b$
and the legs to be sewn together in order to build the $g$ loops with
the indices $2\mu-1$, $2\mu$ ($\mu=1,\ldots,g$). To be more precise we
generate the higher genus surface, by identifying in the tree--level
vertex $g$ pairs of Hilbert spaces: the Hilbert spaces ${\cal
H}_{2\mu-1}$ is identified with ${\cal H}_{2\mu}$ by means of the
projective transformation $P(x_\mu)$ and the twist operator
$\ex{2\pi\ii \e_\m j_0^\m}$.
It is standard to indicate the local coordinates of the insertions by
means of the functions $V_i(z) = \gamma_i(1-z)$ satisfying $V_i(0) =
z_i$, see \eq{tree}. $U_i(z)$ is related to the inverse of $V_i(z)$:
$U_i(z)= \Gamma V_i^{-1}(z)$, where $\Gamma$ is the inversion:
$\Gamma(z) = 1/z$. As in~\cite{DiVecchia:1989id}, a tilde on
$V_{2\mu-1}$ or $U_{2\mu-1}$ indicates the composition of the local
coordinates with the projective transformation generated by the
propagator $P$: $\tilde{V} = V P$. In the following formulae we will
always write explicitly the effect of the twist induced by
$\ex{2\pi\ii \e j_0}$. The result of this manipulation is a
generating vertex with $N_b$ insertions on the $g$--torus which
generalizes\footnote{The reader should keep in mind that in this paper
we are interested only in the ``minimal'' correlation
function~\eq{g-l} with $N_b$ insertions of $b$ field; thus in the
following equation, we can ignore all the external oscillators
$b^{(I)}_n$ $\forall n$ and the $c^{(I)}_n$ for $n\geq \l+1$.  In this
respect our treatment is less general than the one
in~\cite{DiVecchia:1989id}. It is straightforward, although even more
cumbersome, to discuss the twisted vertex $\langle V^{bc}_{N_b;g} |$
in full generality.} Eq.~($\cal C$.5)
\begin{equation}\label{nC5}
\langle{V^{bc}_{N_b;g}}| = \left[ \prod_{I=1}^{N_b} {}_{I}\langle
  q=-Q| \right]\langle{-Q_g}|~  \Delta_b\; \exp\Big\{\sum_{I=1}^{N_b}
  c^{(I)}_\l  e_I \Big\}\; {\cal M} \; |-Q_g \rangle~,
\end{equation}
where
\begin{itemize}
\item $|-Q_g \rangle$ stays for $\prod_{\mu=1}^g |q=-Q
\rangle_{(2\mu)}$,
\item $\Delta_b$ is the fermionic delta--function on the zero-modes of
the internal lines already present at tree--level \eq{tree}
\begin{equation}\label{Delta}
\Delta_b = \prod_{r=1-\l}^{\l-1} \left\{\sum_{s=1-\l}^{\l-1}
\sum_{\m=1}^g \left[E_{rs}(\widetilde{V}_{2\m-1}) \ex{2\pi \ii\e_\m}
b_s^{(2\m)} + \ex{-\ii \pi \l} E_{rs}(V_{2\m})b^{(2\m)}_s\right]
\right\}~,
\end{equation}
\item $E_{ns}(\gamma)$ are the matrices of the representation
of the projective group with weight $\l$~\cite{DiVecchia:1989id},
\item $e_I$ summarizes the coupling between the external lines and the
zero--modes of the internal lines
\beq\label{ei}
e_I = \ex{\ii\pi(1-\l)} \sum_{\m=1}^g 
\sum_{s=1-\l}^{\l-1} \Big[E_{\l s}(U_I\widetilde{V}_{2\m-1})
\ex{2\pi\ii \e_\m} b_s^{(2\m)} +
\ex{-\ii\pi\l} E_{\l s}(U_I V_{2\m})b_{-s}^{(2\m)}\Big]
\eeq
\item ${\cal M}$ is the result of the trace over the non--zero modes
  ($n\geq\l$) of the internal lines, that we perform using coherent
  states. The computation is then reduced to a fermionic Gaussian
  integral and the effect of the twists $\ex{2\pi\ii\e_\mu j_0^\mu}$
  is to add some phases in the various coefficients of the Gaussian
  integral~($\cal C$.9).
\end{itemize}
The result of the trace over the internal non--zero modes can be
written in the usual form
\beq\label{nC13}
{\cal M} = {\rm det} (1-H) \;\exp{\!\left[-(C_1 C_2)\, (1-H)^{-1}
\left(\!\! \begin{tabular}{c} $B_1$ \\ $B_2$
\end{tabular}\!\!
\right) \right]}
\eeq
with a slightly modified definition for the quadratic form $H$ and for
the linear parts $C_1$, $B_1$ and $B_2$ ($C_2$ is unchanged). It is
simple to see that the new $H$ is like~($\cal C$.9) with the first $g$
blocks of columns multiplied by $\ex{2\pi\ii \e_\nu}$ and last $g$
blocks of rows multiplied by $\ex{-2\pi\ii \e_\mu}$
\beq\label{tH}
H^{\mu\nu}_{nm} = \left(
\begin{tabular}{cc}
$E_{nm}(U_{2\mu} \widetilde{V}_{2\nu-1})~ \ex{2\pi\ii\e_\nu}$ &
$\ex{-\pi\ii \l}\; E_{nm}(U_{2\mu}{V}_{2\nu})$ \\
$\ex{-\pi \ii\l}\; E_{nm}(\widetilde{U}_{2\mu-1} \widetilde{V}_{2\nu-1})
~ \ex{2\pi\ii(\e_\nu-\e_\mu)}$ &
$E_{nm}(\widetilde{U}_{2\mu-1} {V}_{2\nu})~ \ex{-2\pi\ii(\e_\mu+\l)}$
\end{tabular}
\right)~,
\eeq
whith $n,m \geq \l$, $\mu,\nu=1,\ldots,g$; moreover in the
off-diagonal entries we used the convention $E_{nm}(U_{2\mu}
V_{2\mu}) = E_{nm}(\widetilde{U}_{2\mu-1} \widetilde{V}_{2\mu-1}) =
0$, $\forall n,m$. 
%
The new linear terms are
\begin{eqnarray}
(C_1)^\mu_m & = & \ex{\pi \ii (1-\lambda)} \sum_{I=1}^N
c_\l^{(I)} E_{\l m}(U_I\widetilde{V}_{2\mu-1})
\ex{2\pi\ii \e_\m}~, \nonumber \\ \label{nC11-C12}
(C_2)^\mu_m & = & - \ex{2\pi\ii\lambda} \sum_{I=1}^N
c_\l^{(I)} E_{\l m}(U_I {V}_{2\mu})
~, \\ \nonumber
(B_1)^\mu_n  & = & - \sum_{\nu=1}^g \sum_{r=1-\lambda}^{\l-1} \left[
E_{nr}(U_{2\m} \widetilde{V}_{2\nu-1}) b_r^{(2\nu)} \ex{2\pi\ii \e_\nu}
+ \ex{- \pi \ii \lambda} E_{nr}(U_{2\m} \widetilde{V}_{2\nu})
b_{-r}^{(2\nu)} \right]
~, \\ \nonumber
(B_2)^\mu_n  & = &  \ex{\pi \ii (1-\lambda)} \ex{-2\pi\ii \e_\mu}
\!\sum_{\nu=1}^g \sum_{r=1-\lambda}^{\l-1}\!\! \left[ \!
E_{nr}(\widetilde{U}_{2\m-1} \widetilde{V}_{2\nu-1}) \ex{2\pi\ii
\e_\nu} b_r^{(2\nu)}  + E_{nr}(\widetilde{U}_{2\m-1} {V}_{2\nu})
b_{-r}^{(2\nu)} \right] .
\end{eqnarray}
Again we have neglected the terms containing $c_n^{(I)}$, $n\geq\l+1$
and all $b^{(I)}$ which are present in the complete expressions of
$(B_1)$ and $(B_2)$.

It is important to notice that the presence of the twists along the
$b$--cycles does not modify the Schottky group structure present
in~\cite{DiVecchia:1989id}: the building blocks of the computation are
still written in terms of the $E^\lambda$ representation of the
projective group, and the twists appear only as a multiplicative phase
of the usual $E^\lambda$ matrices. In particular this means that in
the calculation of $\cal M$ one still reconstructs the $g$ Schottky
generators through the usual combination of local coordinates and the
untwisted propagator
\beq \label{gen}
S_\mu = \widetilde{V}_{2\mu-1} U_{2\mu} = V_{2\m-1} \, P\,
\Gamma \, V^{-1}_{2\m} ~.
\eeq
Because of this reason, it is not difficult to see, by following
appendices E and D of \cite{DiVecchia:1988cy}, that ${\rm det} (1-H)$
is exactly given by Eq.~\eq{tdetH}.

In order to compute the zero--mode contribution ({\em i.e.} the
expectation value in~\eq{nC5}), it is necessary to expand the factor
of $(1-H)^{-1}$ in the exponent of~\eq{nC13} in powers of
$H$. Following the same steps of~\cite{DiVecchia:1989id} one gets
\begin{equation}
\exp\Big\{\sum_{I=1}^{N_b} c^{(I)}_\l e_I \Big\}\; {\cal M} =
{\rm det} (1-H) \;\exp{\!\left[ \sum_{I=1}^{N_b} c^{(I)}_\l f_I
    \right]}~,
\end{equation}
where
\begin{equation}\label{fi}
f_I = \ex{-2i\pi\l} \sum_\a \sum_{\mu=1}^g \sum_{m=\l}^\infty
\sum_{r,s=1-\l}^{\l-1}\, E_{\l m}(U_I T_\a)\, E_{mr}(S_\mu)
E_{rs}(U_\mu)\; \ex{2\pi \ii (\e\cdot N_\a +\e_\mu)}\;
b_{-s}^{(2\mu)}~,
\end{equation}
where the sum $\sum_\a$ is extended over all the elements $T_\a$ of
the Schottky group ${\cal S}_g$. The operator ${\cal F}_{N_b,g}$ in
Eq.~\eq{nC37} is then given by
\begin{equation}\label{ncalfA}
{\cal F}_{N_b,g} = \langle{-Q_g}|\; \Delta_b \; \exp{\!\left[
    \sum_{I=1}^{N_b} c^{(I)}_\l f_I \right]} \;|{-Q_g}\rangle~.
\end{equation}
The explicit evaluation of this equation is particularly simple in the
case $\l=1$, where it is directly related to the twisted abelian
differential~\eq{tda}. First, the sums and the products on
$r,s=1-\l,\ldots,\l-1$ have a single term $r,s=0$; moreover from the
explicit form of the representation $E$, see ($\cal A$.1) of
~\cite{DiVecchia:1989id}, it is easy to see that $E_{00}(\gamma)=1$
for all the projective transformations~$\gamma$. Thus Eq.~\eq{Delta}
simply becomes $\Delta_b = \sum_{\mu=1}^g (\ex{2\pi \ii \e_\mu} -1)
b_0^{(2\mu)}$, while the sum $\sum_m E_{1m}(U_I T_\a) E_{m0}(S_\mu)$
present in $f_I$~\eq{fi} can be treated as done in Appendix~D
of~\cite{DiVecchia:1989id}\footnote{We refer in particular to
Eq.~($\cal D$.2). Notice that here we present a generalization of that
identity, since~\eq{nd2} is valid for all projective transformation
$T_\a$, without the need of the sum over all the element of the
Schottky group $\sum_\a$.}
\bea\label{nd2}
\sum_{m=1}^\infty E_{1m}(U_I T_\a) E_{m0}(S_\mu) & = &
-(T'_\a)^{-1}(z_I) \left[\frac{1}{T_\a^{-1}(z_I) - S_\mu(0)}
-  \frac{1}{T_\a^{-1}(z_I) } \right]
\\ \nonumber &
=& -\left[\frac{1}{z_I - T_\a S_\mu(0)}-\frac{1}{z_I - T_\a(0)} \right]~.
\eea
The first identity follows from Eq.~($\cal A$.11)
of~\cite{DiVecchia:1989id} and from the simplest possible choice of
the local coordinates $V_I(z) = z_I - z$, while the second line is
obtained by using~\eq{ideu}. At this point one can easily compute the
scalar product in the $g$ Hilbert spaces of the loops~\eq{ncalfA} and
in the $g-1$ Hilbert spaces of external fields~\eq{ncalf}, simply by
recalling that $\langle q=1| b_0 | q=1\rangle = 1 $. This means that
one has to select a factor of $b_0$ for each one of the
$\mu=1,\ldots,g$ Hilbert spaces ${\cal H}_\mu$ and one $c^{(I)}_1$ for
each of external Hilbert space ${\cal H}^{(I)}$: this yields the
determinant~\eq{detq1}. Then a slightly non--trivial step is to show
that the constituents of this determinant can be written in the form
of Eq.~\eq{si1}. The basic idea is to single out the sum over the
elements of the form $T_\a S_\mu^\ell$. For sake of notational
simplicity, we use the following abbreviations $x =
\exp(2\pi\ii\e_\mu)$ and $c_\ell = 1/(z-T_\a S_\mu^\ell (0))$, so
Eq.~\eq{Om} can be rewritten as
\beq\label{zrew1}
\zeta_\mu(z_I) = {\sum_\a}^{(\mu)} x \sum_{\ell=-\infty}^{\ell=\infty}
x^{\ell} \left( c_{\ell+1} - c_\ell  \right)~.
\eeq
Let us focus on the series over $\ell$ in the parenthesis and relabel
the summed index in order to combine the two terms of~\eq{zrew1}
\beq\label{zrew2}
\lim_{N_1,N_2 \to\infty} \sum_{\ell=-N_2}^{N_1} x^\ell (c_{\ell + 1} -
c_\ell) =
\lim_{N_1,N_2\to\infty} \left\{(1-x) \sum_{\ell=-N_2+1}^{N_1}c_\ell
x^{\ell -1} + x^{N_1} c_{N_1+1} - x^{-N_2} c_{-N_2}
\right\}~.
\eeq
The convergence of the series is made manifest by breaking the sum
into two pieces ($\ell >0$ and $\ell \leq 0$) and by adding ad
subtracting the limiting values
\beq\label{limv}
A=\lim_{\ell\to\infty} c_\ell = \frac{1}{z-T_\a(\eta_\mu)}~~,~~~
R=\lim_{\ell\to-\infty} c_\ell = \frac{1}{z-T_\a(\xi_\mu)}~.
\eeq
So the parenthesis in~~\eq{zrew1} can be rewritten as follows
\begin{eqnarray}\label{zrew3}
\{\ldots\} & = & \lim_{N_1,N_2\to\infty} \Big\{
(1-x) \sum_{\ell=1}^{N_1} x^{\ell-1} (c_\ell -A) + (1-x^{N_1}) A
+ x^{N_1} c_{N_1+1} +
\\ \nonumber & &~~
(1-x) \sum_{\ell=0}^{N_2-1} x^{-\ell-1} (c_{-\ell} -R) - (1-x^{-N_2}) R
- x^{-N_2} c_{-N_2}\Big\}~.
\end{eqnarray}
Thanks to~\eq{limv} the terms proportional to $x^{N_i}$ cancel in the
limit and one obtains
\beq\label{zrew4}
\{\ldots\} = A - R - (1-x^{-1}) \left[ \sum_{\ell=1}^\infty x^\ell
(c_\ell - A) + \sum_{\ell=0}^\infty x^{-\ell} (c_{-\ell} - R) \right]~.
\eeq
When this result, together with~\eq{limv}, is used in~\eq{zrew1}, the
first two terms give exactly the first line of~\eq{si1}, while the
second line of~\eq{si1} is reproduced by the square parenthesis.

\vspace{.5cm}

In the case $\l=3/2,2,\ldots$ the computations are more cumbersome
because we have $(2\l -1)$ zero--modes and then the number of
insertions necessary to have a non--trivial result is $N_b = (2\l -1)
( g-1)$. Similarly to what has just been done for $\l=1$, one can
start from~\eq{ncalfA} and derive ${\cal F}^{(\l)}$ which is the
generalization of~\eq{ncalf} for a generic spin $\l$. It is
convenient to replace the index $I = 1,\ldots,N_b$ with a double index
$(\mu,r)$ with $\mu=1,\ldots,g-1$ and $r=1-\l,\ldots,\l-1$. With this
notation one gets
\beq\label{detq1l}
{\cal F}^{(\l)}
={\rm det}\left(
\begin{tabular}{ccc}
$\boldsymbol{\zeta}_{1}(\boldsymbol{z}_{1})$ & \ldots &
$\boldsymbol{\zeta}_g(\boldsymbol{z}_1)$ \\
\vdots & & \vdots \\
$\boldsymbol{\zeta}_1(\boldsymbol{z}_{g-1})$ & \ldots &
$\boldsymbol{\zeta}_{g} (\boldsymbol{z}_{g-1})$  \\
$\boldsymbol{\cal E}(S_1)$  & \ldots & $
\boldsymbol{\cal E}(S_g)$
\end{tabular}
\right)~,
\eeq
where ${\cal E}(S_\m)_{rs}$ is defined in~\eq{cale} and
where each entry of~\eq{detq1l} is a $(2\l-1)\times(2\l-1)$ matrix
$[\boldsymbol{\zeta_\mu}(\boldsymbol{z}_\nu)]_{rs} =
\zeta_{\mu,s}(z_{\nu,r})$
\beq\label{zgl}
\zeta_{\mu,r}(z) =\left. \sum_\a \ex{-2\pi\ii (\e\cdot
  N_\a-\e_\mu)} \frac{\partial_y^{(r+\l-1)}}{(r+\l-1)!}
\frac{(T'_\a(z))^\l (S'_\mu(y))^{1-\l}}{T_\a(z) -S_\m(y)}
\left(\frac{S_\m (y)}{T_\a (z)}\right)^{2\l-1}\right|_{y=0}\,.
\eeq
The convergence of the Poincar\'e series which defines $\zeta$ is
assured by the factor $(T'_\a(z))^\l$. In fact it is easy to show that
$T'_\a(z) = k_\a (T_\a(z)-\xi_\a)^2/(z-\xi_\a)^2$. Moreover, the
multiplier of a composite transformation $T_\a$ depends on the
multipliers (and the fixed points) of the generators present in
$T_\a$; in particular $k_\a$ contains a factor $k_\m^l$, if the
generator $S_\m$ or its inverse appear $l$ times in the expression
of $T_\a$. Thus when the order $\a$ increases, the element in the
sum~\eq{zgl} contains high powers of some multiplier, thanks to the
factor $(T'_\a(z))^\l$. Since $|k|<1$ the only condition for the
convergence of the series is $\l>0$.

It is quite easy to check that for $\l=1$ the determinant~\eq{detq1l}
reduces to the result in Eq.~\eq{detq1}. In fact
\bea \nonumber
\zeta_\m(z) dz & = & \sum_\a \ex{-2\pi\ii (\e\cdot N_\a-\e_\mu)}
\frac{S_\m(0) \; d T_\a(z) }{(T_\a(z) -S_\m(0)) T_\a(z)}
\\ \label{lto1} & = &
 \sum_\a \ex{-2\pi\ii (\e\cdot N_\a-\e_\mu)} d \log\frac{T_\a(z)
   -S_\m(0)}{T_\a(z)} \\ \nonumber & = &
 \sum_\a \ex{2\pi\ii (\e\cdot N_\a
+\e_\mu)}\left[\frac{1}{z - T_\a S_\mu(0)}-\frac{1}{z - T_\a(0)}
\right]\,dz~.
\eea
where in the last line we used Eq.~\eq{ideu} and then relabeled
$T_\a^{-1}$ as $T_\a$ in the sum. In the general case
$\l=3/2,2,\ldots$, the analysis is qualitatively similar to the $\l=1$
case described in Section~3. On one hand the $\zeta_{\m,r}$ defined
in~\eq{zgl} have the expected periodicity properties, since they are
single valued when $z$ is moved around an $a$--cycle, while around a
$b_\nu$--cycle they transform as
\beq\label{bctl}
\zeta_{\m,r} (S_\nu (z))\, \left(d S_\nu (z)\right)^\l =
\ex{2\pi\ii\e_\nu} \zeta_{\m,r}(z)\, (dz)^\l~.
\eeq
However, the $\zeta_{\m,r}$ in~\eq{zgl} are too many to be
identified with the twisted $\l$--differentials, since they are
$(2\l-1) g$, instead of $(2\l-1) (g-1)$, as expected. In fact this
identification is not possible, since the $\zeta_{\m,r}$ are not
holomorphic, because of the pole singularity in $z=0$. The singular
part comes only from two elements of the Poincar\'e series~\eq{zgl}:
$T_\a=I$ and $T_\a=S_\m$. So it is easy to extract this singularity
for $z\to 0$
\beq\label{spl}
\zeta_{\m,r}(z) \sim - \sum_{s=1-\l}^{\l-1} \frac{1}{z^{s+\l}} {\cal
  E}(S_\m)_{sr}~.
\eeq
Therefore one can build $(2\l-1) (g-1)$ twisted holomorphic
differentials by generalizing Eq.~\eq{tda}
\beq\label{tdal}
\Omega_{\mu,r} (z) = \left[\zeta_{\mu,r} (z) -  \sum_{s,t=1-\l}^{\l-1}
\zeta_{g,s} (z)\, \left({\cal E}(S_g)\right)^{-1}_{st} {\cal
  E}(S_\m)_{tr} \right]~~,\quad \mu=1,\ldots,g-1~.
\eeq
In order to write the $\Omega$'s we have of course supposed that
${\cal E}(S_g)$ is invertible which is surely true for any $\l$ if
$\e_g \not= 0$. In fact one can easily diagonalize the $(2l-1)\times
(2\l-1)$ matrix ${\cal E}(S_g)$ and check that its eigenvalues are
\beq\label{evl}
\left(\ex{2\pi\ii\e_g} k_g^r -1 \right)~~,~~~~\mbox{with}~~
r=1-\l,\ldots,\l-1 ~,
\eeq
and none of them is vanishing. As for $\l=1$~\eq{detq2}, one can
write~\eq{detq1l} in terms of these $\l$--differentials
\beq\label{detq2l}
{\cal F}^{(\l)} ={\rm det}\left(
\begin{tabular}{cccc}
$\boldsymbol{\Omega}_{1} (\boldsymbol{z}_{1})$ & \ldots &
$\boldsymbol{\Omega}_{g-1}(\boldsymbol{z}_{1})$ &
$\boldsymbol{\zeta}_{g}(\boldsymbol{z}_{1})$
  \\  \vdots & & \vdots & \vdots\\
$\boldsymbol{\Omega}_{1}(\boldsymbol{z}_{g-1})$ & \ldots &
$\boldsymbol{\Omega}_{g-1}(\boldsymbol{z}_{g-1})$ &
$\boldsymbol{\zeta}_{g}(\boldsymbol{z}_{g-1})$
 \\  $0$  & \ldots & $0$ &
$\boldsymbol{\cal E}(S_g) $
\end{tabular}
\right)~,
\eeq
and the final result for the ``minimal'' correlator we are interested
in is given by the product of~\eq{tdetH} and \eq{detq2l}: $Z^\l_\e =
{\rm det} (1-H)\, {\cal F}^{(\l)}$.

For half--integer $\l$ ($\l=3/2,5/2,\ldots$), the same computation can
be done also for the case where $\e_\m = 0$ $\forall \m$. In fact, all
the matrices ${\cal E}(S_\m)$ are still invertible also in this
case. This can be easily seen from~\eq{evl} which tells that for
half--integer $r$ the eigenvalues of any ${\cal E}(S)$ are
non--vanishing even if $\e=0$.  In order to compare with the result
of~\cite{DiVecchia:1989id}, one can show that the differential
$\Lambda_\m(z)$ introduced there, see Eq.~(${\cal D}$.12), can be
derived also starting from the $\zeta_\m$ defined in~\eq{zgl} setting
$\e_\m=0~\forall \m$
\beq\label{Labc}
\Lambda_\m(z) = - \zeta_\m\, {\cal E}(S_\m)^{-1} +
\zeta_g\, {\cal E}(S_g)^{-1} ~.
\eeq
This means that for  half--integer $\l$ the twists along the $b$-cycles
do not introduce any qualitative change in the geometric
interpretation of the final result and their effect is just to bring
some phase factor in the explicit definition of the
differentials. This is to be contrasted with the $\l=1$ case where the
counting of the zero--modes of the $\bar\partial$ operator changes
in presence of non--trivial twists.

The case $\l=1/2$ is
special~\cite{Pezzella:1988jr,Losev:1989fe,DiVecchia:ht} since
all the zero--modes are absent. Of course one can still write a
Poincar\'e series similar to~\eq{zgl}
\beq\label{sk}
\sum_\a \ex{-2\pi\ii (\e\cdot N_\a-\e_\mu)} \frac{(T'_\a(z))^{1/2}
(S'_\mu(0))^{1/2}}{T_\a(z)  -S_\m(0)}\,,
\eeq
which is an automorphic form of weight $1/2$ closely related to the
Sz\"ego kernel. However it is not possible to use the trick
in~\eq{Labc} and eliminate the pole in $z=0$ in order obtain some
non--trivial holomorphic differentials. In fact, the substitution
$T_\a=S_\m T_\beta$ shows that~\eq{sk} does not depend on $\mu$ and
thus subtracting two values of $\m$ in~\eq{sk} , as done in~\eq{Labc},
one obtains identically zero.

For integer $\l\geq 2$ and vanishing twists $\e=0$, we have to
distinguish two cases. On the torus $g=1$ both $b$ and $c$ have a
zero--mode, see Eq.~\eq{zm1l}. On higher genus surfaces $g\geq 2$,
only $b$ has zero-modes. Thus, for higher genus surfaces, the $\e=0$
case is conceptually on the same footing as the twisted case and one
can build $(2\l-1) (g-1)$ differentials starting from~\eq{zgl}. First
it is easy to see why the zero--modes of $c$ are absent in this
case. Usually they are constructed starting from the $\zeta_{\m,r}$
in~\eq{zgl} by sending $\l \to 1-\l$. Formally this satisfies the
periodicity condition, but the series does not converge since the
factor of $T'_\a(z)$ has now a negative power. Since there are no
zero--modes for $c$, the determinant~\eq{detq1l} does not vanish even
if ${\rm det}[{\cal E}(S_\m)]=0$ $\forall \m$ when $\e=0$. Hence the
Riemann--Roch theorem for $g\geq 2$ is realized in the same way,
independently of possible twists for $\l =3/2,2,\ldots$.

\end{document}